\documentclass[11pt]{article}
\usepackage{amsmath,amsthm,amssymb,hyperref, color,fullpage,bm}
\usepackage{blkarray}

\newcommand{\remove}[1]{}

\makeatletter
\newtheorem*{rep@theorem}{\rep@title}
\newcommand{\newreptheorem}[2]{%
\newenvironment{rep#1}[1]{%
 \def\rep@title{#2 \ref{##1}}%
 \begin{rep@theorem}}%
 {\end{rep@theorem}}}
\makeatother

\hypersetup{
	colorlinks=true,
	linkcolor=blue,%
	citecolor=blue,
	linktoc=page
}

\newtheorem{thm}{Theorem}[section]
\newreptheorem{thm}{Theorem}

\newtheorem{lem}[thm]{Lemma}
\newtheorem{define}[thm]{Definition}

\newtheorem{prop}[thm]{Proposition}

\def\F{{\mathbb{F}}}

\def\bbeta{{\bm \beta}}

\def\_{\,\,\,\,\,}

\usepackage{leftidx}

\usepackage{tikz-cd}

\begin{document}

\title{A construction of Maximally Recoverable LRCs for small number of local groups}

\date{}

\author{Manik Dhar\thanks{Department of Computer Science, Princeton University. Email: \texttt{manikd@princeton.edu}. Part of this work was done while this author was working at Microsoft research. Research supported by NSF grant DMS-1953807.} \and
Sivakanth Gopi\thanks{Microsoft Research.
Email: \texttt{sigopi@microsoft.com}.}}
\maketitle

\begin{abstract}
Maximally Recoverable Local Reconstruction Codes (LRCs) are codes designed for distributed storage to provide maximum resilience to failures for a given amount of storage redundancy and locality. An $(n,r,h,a,g)$-MR LRC has $n$ coordinates divided into $g$ local groups of size $r=n/g$, where each local group has `$a$' local parity checks and there are an additional `$h$' global parity checks. Such a code can correct `$a$' erasures in each local group and any $h$ additional erasures. Constructions of MR LRCs over small fields is desirable since field size determines the encoding and decoding efficiency in practice. In this work, we give a new construction of $(n,r,h,a,g)$-MR-LRCs over fields of size $q=O(n)^{h+(g-1)a-\lceil h/g\rceil}$ which generalizes a construction of Hu and Yekhanin (ISIT 2016). This improves upon state of the art when there are a small number of local groups, which is true in practical deployments of MR LRCs.
\end{abstract}


\section{Introduction}
In modern distributed storage systems, data is split and stored in individual servers. A server crash can potentially lead to the loss of all data in a server. Even if a server becomes temporarily unavailable, for instance due to serving too many requests, that could lead to very slow access to data on that server.  Replication of data is a potential solution but that is very inefficient in terms of storage. Error correcting codes offer a more efficient solution. For example, distributed storage systems such as RAID use Reed-Solomon codes. An $(n,k)$-Reed Solomon code will add $n-k$ redundant servers (parity checks) to $k$ data servers and allows us to recover from an arbitrary $n-k$ erasures\textbf{} by reading the remaining $k$ servers. But for large $k$, which is needed to get good storage efficiency, this would require us to read a lot of servers to recover lost data. 

Local reconstruction codes (LRCs) were invented to deal with this problem. \emph{Locality} means that when a small number of servers fail, any failed server can be recovered quickly by reading data from a small number of healthy servers. At the same time they can recover from catastrophic failures where a large number of servers fail (although recovery will necessarily be less efficient). Locality in distributed storage was first introduced in~\cite{huang_erasure_2012,Chen2007OnTM}, but LRCs were first formally defined and studied in \cite{gopalan_locality_2012} and \cite{Papailiopoulos2012LocallyRC}.

\begin{define}
An $(n,r,h,a,g)$-LRC is a linear code $C$ over $\F_q$ of length $n$ (in other words a subspace $C\subseteq \F_q^n$), whose codeword symbols are partitioned into $g$ local groups each of size $r=n/g$. The coordinates in each local group satisfy `$a$' local parity checks (in other if the codewords are column matrices then the co-ordinates in a group lie in the kernel of an $a\times r$ matrix) and there are further $h$ global parity checks that all the $n$ coordinates satisfy (in other words the codewords lie in the kernel of an $h\times n$ matrix). 

The local parity checks are used to recover from up to $a$ erasures in a local group by reading at most $r-a$ symbols in that local group. The $h$ global parities are used to correct more global erasure patterns which involve more than $a$ erasures in each local group.

\end{define}
 
 The above definition implies that the parity check matrix $H$ of an $(n,r,h,a,g)$-LRC has the structure shown in Equation~\ref{eq-MR}.

\begin{equation}\label{eq-MR}
H=
\left[
\begin{array}{c|c|c|c}
A_1 & 0 & \cdots & 0\\
\hline
0 &A_2 & \cdots & 0\\
\hline
\vdots & \vdots & \ddots & \vdots \\
\hline
0 & 0 & \cdots & A_g \\
\hline
B_1 & B_2 & \cdots & B_g \\
\end{array}
\right]
\end{equation}

Recall that $g$ is the number of local groups and each local group has size $r=n/g$. $A_1,A_2,\dots,A_g$ are $a\times r$ matrices over $\F_q$ which correspond to the local parity checks that each local group satisfies. $B_1,B_2,\dots,B_g$ are $h\times r$ matrices over $\F_q$ and together they represent the $h$ global parity checks that the codewords should satisfy.

We are interested in LRCs which can correct as many erasures as possible. The best one could hope for is a set of values for entries in $H$ (as shown in \eqref{eq-MR}) which can correct any set of erasures a generic matrix of that form can (one way to define a generic $H$ is to assume every entry is algebraically independent of each other). We know that a set of erasures $E$ can be recovered from if and only if the subset of columns of $H$ corresponding to $E$ have full rank. Using this it can be shown that a generic $H$ can correct an additional $h$ erasures distributed across local groups on top of the `$a$' erasures in each local group. LRCs which can correct all such erasure patterns are called \emph{Maximally Recoverable (MR) LRCs}. Maximal recoverability was first introduced by~\cite{Chen2007OnTM,huang_erasure_2012} and extended to more general settings in~\cite{gopalan_explicit_2014}. MR-LRCs were also studied earlier in~\cite{Blaum2012PartialMDSCA} where they are called \emph{Partial-MDS (Maximum Distance Separable) codes}.

For clarity, we formally define MR-LRCs.

\begin{define}
Let $C$ be an arbitrary $(n,r,h,a,g)$-LRC where $r=n/g$. We say that $C$ is maximally recoverable if:
\begin{enumerate}
	\item Any set of `$a$' erasures in a local group can be corrected by reading the rest of the $r-a$ symbols in that local group.
	\item Any erasure pattern $E\subseteq [n],$ $|E|=ga+h,$ where $E$ is obtained by selecting $a$ symbols from each of $g$ local groups and $h$ additional symbols arbitrarily, is correctable by the code $C.$
\end{enumerate}
\end{define}

Again, our discussion gives us the following characterization for the parity check matrices of MR-LRCs.
\begin{prop}
	\label{prop-parity}
	An $(n,r,h,a,g)$-LRC with parity check matrix given by $H$ from Equation~\ref{eq-MR} is maximally recoverable iff: 
\begin{enumerate}
	\item Each of the local parity check matrices $A_i$ are the parity check matrices of an MDS code, i.e., any $a$ columns of $A_i$ are linearly independent.
	\item Any submatrix of $H$ which can be formed by selecting $a$ columns in each local group and additional $h$ columns has full column rank.
\end{enumerate}
\end{prop}

Practical deployments of MR LRCs typically have a small number of local groups (say $g=2,3,4$) and a small number of local parities per local group (say $a=1,2$)~\cite{huang_erasure_2012}. Moreover, field size is the most important determinant of encoding and decoding efficiency of MR LRCs~\cite{huang_erasure_2012}, since encoding and decoding requires several finite field operations. Therefore constructions of MR LRCs over small fields in this regime are important in practice. 

In this work, we give a new construction of MR LRC tailored for this regime. 
\begin{thm}\label{thm-mrsk}
There exists an explicit $(n,r,h,a,g)$-MR-LRC over a field of size $O(n)^{h+(g-1)a-\lceil h/g\rceil}$.
\end{thm}

For the $a>1$ and small $g$ parameter regime, the general construction of \cite{gopi_improved_2022} shown in (\ref{eq:GG22_bound}) is the only one we need to compare to. Here we see that for small constant $g$, which means $r=\Omega(n)$, as long as $h-\lceil h/g\rceil+a(g-1)<\min\{h,r-a\}$ our construction gives us a better MR-LRC construction. In particular, when $g=2$, $h$ is even, and $a$ is a constant our construction is over fields of size $O(n)^{h/2+a}$ as opposed to $O(n)^{h}$ from \cite{gopi_improved_2022}.

Our construction generalizes a construction from~\cite{hu_new_2016} which is specialized for $a=1$ and requires a field size of $n^{h-\lceil h/g\rceil +1}$. While the proof in \cite{hu_new_2016} can also be generalized, this paper analyses the generalization with a new and direct proof using simple linear algebra and basic properties of Vandermonde matrices and Gabidulin codes. This also gives a new proof for the original result of \cite{hu_new_2016}. Our construction and its proof is inspired by the constructions of MR LRCs in~\cite{gopi_improved_2022,cai_construction_2021}.  

\subsection{Prior Work}

\paragraph{Upper Bounds:} There are a several construction of MR LRCs since their introduction~\cite{blaum_construction_2016,barg_construction_2022,gopalan_explicit_2014,martinez-penas_universal_2019,guruswami_constructions_2020,gabrys_constructions_2018,martinez-penas_general_2022,liu_maximally_2022,gopi_maximally_2020,chen_sector-disk_2015}. The current best constructions of MR-LRCs over most range of parameters is due to \cite{gopi_improved_2022,cai_construction_2021} which require a field size of
\begin{equation}
\label{eq:GG22_bound}
    \left(O\bigl(\max\{n/r,r\}\bigr)\right)^{\min\{h,r-a\}}.
\end{equation}
These codes are constructed using the theory of skew-symmetric polynomials. As mentioned earlier our construction provides better field sizes in the case of constant $g$. For $g$ constant, constant $a$, and $h$ divisible by $g$ we have $r=\Omega(n)$ which means our construction gives a field size of $O(n)^{h(1-1/g)+a}$ and \ref{eq:GG22_bound} gives $O(n)^h$. Our construction does better as long as $a<h/g$.
For more constrained settings much better bounds are known as shown in Table~\ref{tab:upper_bounds}.
\renewcommand{\arraystretch}{1.5}
\begin{table}[!ht]
\caption{Table showing the best known upper bounds on the field size of $(n,r,h,a,g)$-MR LRCs over several constrained settings.}
\label{tab:upper_bounds}
\centering
\begin{tabular}{ |c|c|c|}
\hline
$O(r)$ when $h=0$ or $h=1$& \cite{Blaum2012PartialMDSCA}\\
\hline
$O(n)^{h-\lceil h/g\rceil+g-1}$ when $a=1$ & \cite{hu_new_2016} \\
\hline
$O(n)^{h/2}$ when $a=1,g=2$ and $h \text{ (mod }4\text{)}=0$ & \cite{hu_new_2016} \\
\hline
$O(n)$ when $h=2$ & \cite{gopi_maximally_2020}\\
\hline
$O(n^3)$ when $h=3$ & \cite{gopi_maximally_2020}\\
\hline
$\widetilde{O}(n)$ when $h=3,a=1,r=3$ & \cite{gopi_maximally_2020}\\
\hline
$\bigl(O(n)\bigr)^{\lceil\min\{h,r-1\}(1-1/q_0)\rceil}$ when $a=1$ and $q_0\ge g+1$ is a prime power & \cite{gopi_improved_2022}\\
\hline
\end{tabular}
\end{table}
\renewcommand{\arraystretch}{1}

When $g$ and $r$ are powers of a prime $p$ and $h \text{ (mod }p\text{)}\ne 1$ and $\lceil h/g\rceil \text{ (mod }p\text{)}\ne p-1$ then \cite{hu_new_2016} can shave off a factor of $n$ to give $(n,r,h,a=1,g)$-MR LRCs over fields of size $O(n)^{h-\lceil h/g\rceil+g-2}$ (when $p=g=2$ this gives the third row of our table).  

\paragraph{Lower bounds:} The best known lower bounds on the field size required for $(n,r,h,a,g)$-MR LRCs (with $gr=n$) is from~\cite{gopi_maximally_2020} who show that for $h\ge 2$,
\begin{equation}\label{eqn-lowerbound}
q \ge \Omega_{h,a} \left( n\cdot r^\alpha \right) \text{ where } \alpha=\frac{\min\left\{a,h-2\lceil h/g\rceil\right\}}{\lceil h/g \rceil}.
\end{equation}
When $g\ge h$, this simplifies to $q \ge \Omega_{h,a} n\cdot r^{\min\{a,h-2\}}$.
If $g<h$, in particular when $g$ divides $h$, then the lower bound simplifies to,
\begin{equation}
	\label{eqn:lowerbound_smallg}
q\ge \Omega_{h,a}\left(nr^{\min\{ag/h,g-2\}}\right)
\end{equation}
If in particular we look at $g=2$, we see that the current lower bound is linear while the current best constructions for $a=1$ and $h$ divisible by $4$ is $n^{h/2}$~\cite{hu_new_2016}. This shows that there is a large scope of improvement in either direction.
Any progress in this question is very interesting because as mentioned earlier the regime of small number of groups is important in practice for distributed storage~\cite{huang_erasure_2012}.





\section{Proof of Theorem~\ref{thm-mrsk}}

We are going to use a simple property of Moore matrices which are used to construct Gabidulin codes~\cite{gabidulin,gabidulin_rank_2021}.
 \begin{lem}\label{lem:gabidulin}
Let $\bbeta_1,\dots,\bbeta_n\in \F_{q^m}$ be linearly independent over $\F_q$ (this requires $n\le m$). Then the following Moore matrix $M$ has full column rank.
	 $$
	 M =
	 \begin{bmatrix}
	 	\bbeta_1 & \bbeta_2 & \dots & \bbeta_n\\
	 	\bbeta_1^{q}& \bbeta_2^{q}& \cdots & \bbeta_n^{q}\\
	 	\bbeta_1^{q^2}& \bbeta_2^{q^2}& \cdots & \bbeta_n^{q^2}\\
	 	\vdots & \vdots &  &\vdots\\
	 	\bbeta_1^{q^{m-1}}& \bbeta_2^{q^{m-1}}& \cdots & \bbeta_n^{q^{m-1}}
	 \end{bmatrix}
	 $$
 \end{lem}
 We will also need the concept of Schur complement.
 \begin{lem}
     \label{lem:schur}
     Consider the square matrix $$M=\left[\begin{array}{c|c}
          A& B \\
          \hline
          C& D
     \end{array}\right]$$ where $A,D$ are square matrices and $A$ is invertible. Then doing column operations to remove the columns of $B$ using the columns of $A$ will result in the following matrix\footnote{The matrix $D-CA^{-1}B$ is called Schur complement.} $$M'=\left[\begin{array}{c|c}
          A& 0 \\
          \hline
          C& D-CA^{-1}B
     \end{array}\right].$$ 
 \end{lem}
In the remainder of this section, we present the proof of Theorem~\ref{thm-mrsk} using simple linear algebra and the above lemma.
Let $t=a+\lceil h/g\rceil$ and $m=h+ga-t.$ Let $q\ge n$ be a prime power. We will construct a parity check matrix over the field $\F_{q^m}.$ This means our field size will be $q^m=q^{h+ga-t}=q^{h+(g-1)a-\lceil h/g\rceil}$ as desired.

We partition $\F_q$ into $g$ sets $\{x_{i,1},\hdots,x_{i,r}\}$ of size $r$ and some left over elements which we ignore.

Recall that we want to construct a parity check matrix of the form
\begin{equation}\label{eq-MR-2}
H=
\left[
\begin{array}{c|c|c|c}
A_1 & 0 & \cdots & 0\\
\hline
0 &A_2 & \cdots & 0\\
\hline
\vdots & \vdots & \ddots & \vdots \\
\hline
0 & 0 & \cdots & A_g \\
\hline
B_1 & B_2 & \cdots & B_g \\
\end{array}
\right]
\end{equation}
where each $A_i$ is an $a\times r$ matrix and each $B_i$ is an $h\times r$ matrix.
Define $A_i$ as
	 $$
	 A_i =
	 \begin{bmatrix}
  	 	1 & 1 & \dots & 1\\
	 	x_{i,1} & x_{i,2} & \dots & x_{i,r}\\
	      x_{i,1}^2 & x_{i,2}^2 & \dots & x_{i,r}^2\\
	 	\vdots & \vdots &  &\vdots\\
	      x_{i,1}^{a-1} & x_{i,2}^{a-1} & \dots & x_{i,r}^{a-1}
	 \end{bmatrix}.
	 $$
Define $\bbeta_{i,j}\in \F_{q^m}$ for $i\in [g],j\in[r]$ as
	 $$
	 \bbeta_{i,j} =
	 \begin{bmatrix}
	 	x_{i,1}^t\\
	      x_{i,1}^{t+1}\\
	 	\vdots \\
	      x_{i,1}^{t+m-1}
	 \end{bmatrix},
	 $$
  where we are expressing $\bbeta_{i,j}$ in some basis for $\F_{q^m}$ (which is a $\F_q$-vector space of dimension $m$). We now define $B_i$ as
  	 $$
	 B_i =
	 \begin{bmatrix}
	 	x_{i,1}^a & x_{i,2}^a & \dots & x_{i,r}^a\\
	      x_{i,1}^{a+1} & x_{i,2}^{a+1} & \dots & x_{i,r}^{a+1}\\
	 	\vdots & \vdots &  &\vdots\\
	      x_{i,1}^{t-1} & x_{i,2}^{t-1} & \dots & x_{i,r}^{t-1}\\
            \bbeta_{i,1} & \bbeta_{i,2} & \dots & \bbeta_{i,r}\\
            \bbeta_{i,1}^q & \bbeta_{i,2}^q & \dots & \bbeta_{i,r}^q\\
            \vdots & \vdots &  &\vdots\\
            \bbeta_{i,1}^{q^{h+a-t-1}} & \bbeta_{i,2}^{q^{h+a-t-1}} & \dots & \bbeta_{i,r}^{q^{h+a-t-1}}
	 \end{bmatrix}.
	 $$

For convenience we collect the first $t-a$ rows of $B_i$ into $V_{i}$ and the remaining rows into $G_{i}$ so that \begin{equation}
B_i=
\left[
\begin{array}{c}
V_i\\
\hline
G_i
\end{array}
\right].
\end{equation} 
We see that $V_{i}$ contains the `Vandermonde'-like rows and $G_{i}$ contains the `Gabidulin'-like rows. Also note that the powers of $x_{i,j}$ are increasing steadily from $0$ to $t+m-1$ (which is equal to $ag+h-1$); first along $A_i$, then $V_i$ and then the first row of $G_i$ (when expressed in $\F_q$ basis). This is the most crucial part of the construction as we will shortly see. The MR LRC constructions from~\cite{gopi_improved_2022,cai_construction_2021} also use a similar trick.

Given a matrix $M$ and a subset $I$ of its columns, we let $M(I)$ refer to the sub-matrix of $M$ corresponding to the columns $I$. We also use $\bbeta_i$ to denote the $m\times r$ matrix formed by $\bbeta_{i,1},\dots,\bbeta_{i,r}$,
$$\bbeta_i = \begin{bmatrix}
    \bbeta_{i,1} & \bbeta_{i,2} & \dots & \bbeta_{i,r}
\end{bmatrix}.$$

Let $E$ be an erasure pattern of size $ag+h$ formed by selecting $a$ columns in each of the local groups and additional $h$ columns from anywhere in $H$ as shown in (\ref{eq-MR-2}). We want to show that $H(E)$ (which is an $(ag+h)\times (ag+h)$ matrix) is full rank. Showing $H(E)$ is full rank for every correctable erasure pattern $E$ will prove our theorem.

One of the local groups will contain at least $t=a+\lceil h/g\rceil$ many columns. Without loss of generality let us say that is group $g$. We arbitrarily split the columns in group $g$ into $X$ which is of size $t$ and $Y$ which contains the remaining columns. 
For groups $1,\hdots,g-1$ we arbitrarily split the columns selected in each group to a set $S_i$ of size $a$ and $T_i$ which contains the remaining elements chosen.
So we can write $H(E)$ as
	\begin{align*}
	H(E)=
	\left[
	\begin{array}{c|c|c|c}
	A_1(S_1\cup T_1) & \cdots & 0  & 0\\
	\hline
	\vdots & \ddots & \vdots & \vdots \\
	\hline
         0 & \cdots & A_{g-1}(S_{g-1} \cup T_{g-1}) & 0 \\
         \hline
	0 & \cdots & 0 & A_g(X \cup Y) \\
	\hline
	V_1(S_1\cup T_1) & \cdots & V_{g-1}(S_{g-1}\cup T_{g-1}) & V_g(X \cup Y) \\
\hline
	G_1(S_1\cup T_1) & \cdots & G_{g-1}(S_{g-1}\cup T_{g-1}) & G_g(X \cup Y)
 \end{array}
	\right].
	\end{align*}

Now we note that $A_1(S_1),A_2(S_2),\cdots, A_{g-1}(S_{g-1})$ are $a\times a$ matrices of full rank. We now do column operations on $H(E)$, where in the first $g-1$ local groups we use the columns of $A_i(S_i)$ for $i\le g-1$ to remove the columns of $A_i(T_i)$. This results in the lower block $\left[\frac{V_i(T_i)}{G_i(T_i)}\right]$ to change into a Schur complement as follows:
	\begin{align*}
		\left[
		\begin{array}{c|c}
			A_i(S_i) & A_i(T_i)\\
			\hline
   		V_i(S_i) & V_i(T_i)\\
			\hline
			G_i(S_i) & G_i(T_i)\\ 
		\end{array}
		\right] \rightarrow 
		\left[
		\begin{array}{c|c}
			A_i(S_i) & 0\\
			\hline
      		V_i(S_i) & V_i(T_i)-V_i(S_i)A_i(S_i)^{-1}A_i(T_i)\\
        \hline
			G_i(S_i) & G_i(T_i)-G_i(S_i)A_i(S_i)^{-1}A_i(T_i)\\ 
		\end{array}
		\right],
	\end{align*}
for all $i\le g-1$. For convenience we let $G'_i(T_i)=G_i(T_i)-G_i(S_i)A_i(S_i)^{-1}A_i(T_i)$ and $V'_i(T_i)=V_i(T_i)-V_i(S_i)A_i(S_i)^{-1}A_i(T_i)$ for $i\le g-1$. 

At the end of this $H(E)$ transforms into the following:

\begin{align*}
	M_1=
	\left[
	\begin{array}{c|c|c|c|c|c}
	A_1(S_1) & 0 & \cdots & 0&0  & 0\\
	\hline
	\vdots &\vdots & \ddots & \vdots & \vdots & \vdots  \\
	\hline
         0 & 0 & \cdots & A_{g-1}(S_{g-1}) &0 & 0  \\
         \hline
	0 & 0 & \cdots & 0 &0 & A_g(X\cup Y) \\
	\hline
	V_1(S_1) & V'_1(T_1) & \cdots & V_{g-1}(S_{g-1}) & V'_{g-1}(T_{g-1})& V_g(X\cup Y) \\
\hline
	G_1(S_1) & G'_1(T_1) & \cdots & G_{g-1}(S_{g-1}) & G'_{g-1}(T_{g-1}) & G_g(X\cup Y)
 \end{array}
	\right].
	\end{align*} 
It is clear that it suffices to show the following $(h+a)\times (h+a)$ sub-matrix of $M_1$ is full rank.

\begin{align*}
	M_2=
	\left[
	\begin{array}{c|c|c|c|c}
	 0 & \cdots & 0 & A_g(Y) &A_g(X) \\
	\hline
	 V'_1(T_1) & \cdots & V'_{g-1}(T_{g-1})& V_g(Y) & V_g(X) \\
\hline
	G'_1(T_1) & \cdots & G'_{g-1}(T_{g-1}) & G_g(Y) & G_g(X)\\
 \end{array}
	\right].
\end{align*} 

In $M_2$ we look at the sub-matrix $W(X)=\left[\frac{A_g(X)}{V_g(X)}\right]$ (we also let $W(Y)=\left[\frac{A_g(Y)}{V_g(Y)}\right]$). By construction $W(X)$ is a $t\times t$ Vandermonde matrix of full rank. We use $W(X)$ to remove all the remaining columns in $M_3$ corresponding to the rows of $W(X)$ by doing column column operations giving us the following matrix:

\begin{align*}
	M_3=
	\left[
	\begin{array}{c|c|c|c|c}
	 0 & \cdots & 0 & 0& A_g(X) \\
	\hline
	 0 & \cdots & 0& 0 & V_g(X) \\
\hline
	 G''_1(T_1)& \cdots & G''_{g-1}(T_{g-1}) & G_g(Y)-G_g(X)W(X)^{-1}W(Y) &G_g(X)
 \end{array}
	\right],
\end{align*} 
where $G''_i(T_i)=G'_i(T_i)-G_g(X)W(X)^{-1}\left[\frac{0}{V'_i(T_i)}\right]$ for $i\le g-1$. For convenience, we also let $V''_i(T_i)=\left[\frac{0}{V'_i(T_i)}\right]$.

It is now clear that it suffices to show the following $(h+a-t)\times (h+a-t)$ sub-matrix of $M_3$ is full rank.

\begin{align*}
	M_4=
	\left[
	\begin{array}{c|c|c|c}
	 G''_1(T_1)& \cdots & G''_{g-1}(T_{g-1}) & G_g(Y)-G_g(X)W(X)^{-1}W(Y)
 \end{array}
	\right],
\end{align*}

Note that all the entries in $W(X),W(Y),V'(T_i),A_i(S_i),$ and $A_i(T_i)$ are in the base field $\F_{q}$. Column operations on $G_i$ with $\F_{q}$ coefficients retain its structure with $\bbeta$'s replaced by their corresponding $\F_{q}$-linear combinations. Therefore by Lemma~\ref{lem:gabidulin}, it is enough to show that the following $m\times (h+a-t)$ matrix has full column rank\footnote{Note that $\sum_{i=1}^{g-1}|T_i| + |Y| = h+a-t.$}
$$M_5= \left[
	\begin{array}{c|c|c|c}
	 \bbeta''_1(T_1)& \cdots & \bbeta''_{g-1}(T_{g-1}) & \bbeta_g(Y)-\bbeta_g(X)W(X)^{-1}W(Y)
 \end{array}
	\right],$$
where $\bbeta''_i(T_i)=\bbeta_1(T_i)-\bbeta_1(S_i)A_i(S_i)^{-1}A_i(T_i)-\bbeta_g(X)W(X)^{-1}V''_i(T_i)$ for $i\le g-1$.
Now consider the following $(h+ga)\times (h+ga)$ matrix $F$,
$$F= \left[
	\begin{array}{c|c|c|c|c|c|c}
	 A_1(S_1)&A_1(T_1)& \cdots & A_{g-1}(S_{g-1})&A_{g-1}(T_{g-1}) & A_g(Y) & A_g(X) \\
  \hline
  V_1(S_1)&V_1(T_1)& \cdots & V_{g-1}(S_{g-1})&V_{g-1}(T_{g-1}) & V_g(Y) & V_g(X) \\
  \hline
  \bbeta_1(S_1)&\bbeta_1(T_1)& \cdots & \bbeta_{g-1}(S_{g-1})&\bbeta_{g-1}(T_{g-1}) & \bbeta_g(Y) & \bbeta_g(X)
 \end{array}
	\right].$$
By construction, $F$ is a Vandermonde matrix over $\F_q$ and therefore is of full rank. We now repeat column operations on $F$ analogous to the ones we did on $H(E)$. Concretely, we first use the $a\times a$ invertible matrix $A_i(S_i)$ to clear out the columns $A_i(T_i)$ for $i\le g-1$. This gives us the matrix,
$$F_1= \left[
	\begin{array}{c|c|c|c|c|c|c}
	 A_1(S_1)&0& \cdots & A_{g-1}(S_{g-1})&0 & A_g(Y) & A_g(X) \\
  \hline
  V_1(S_1)&V'_1(T_1)& \cdots & V_{g-1}(S_{g-1})&V'_{g-1}(T_{g-1}) & V_g(Y) & V_g(X) \\
  \hline
  \bbeta_1(S_1)&\bbeta'_1(T_1)& \cdots & \bbeta_{g-1}(S_{g-1})&\bbeta'_{g-1}(T_{g-1}) & \bbeta_g(Y) & \bbeta_g(X)
 \end{array}
	\right]$$
 where $\bbeta'_i(T_i)=\bbeta_1(T_i)-\bbeta_1(S_i)A_i(S_i)^{-1}A_i(T_i)$ for $i\le g-1$.
 Next we use the sub-matrix $W(X)=\left[\frac{A_g(X)}{V_g(X)}\right]$ to clear out the columns $W(Y)=\left[\frac{A_g(Y)}{V_g(Y)}\right]$ and $V''_i(T_i)=\left[\frac{0}{V'_i(T_i)}\right]$. This gives us the matrix,
 $$F_2= \left[
	\begin{array}{c|c|c|c|c|c|c}
	 A_1(S_1)&0& \cdots & A_{g-1}(S_{g-1})&0 & 0 & A_g(X) \\
  \hline
  V_1(S_1)&0& \cdots & V_{g-1}(S_{g-1})&0 & 0 & V_g(X) \\
  \hline
  \bbeta_1(S_1)&\bbeta''_1(T_1)& \cdots & \bbeta_{g-1}(S_{g-1})&\bbeta''_{g-1}(T_{g-1})& \bbeta_g(Y)-\bbeta_g(X)W(X)^{-1}W(Y) & \bbeta_g(X)
 \end{array}
	\right].$$
As $F_2$ is a full rank square matrix, $M_5$ should have full column rank which implies $H(E)$ has full rank completing the proof.

\bibliographystyle{alpha}
\bibliography{bibliography.bib}

\newcommand{\etalchar}[1]{$^{#1}$}
\begin{thebibliography}{CMST21}

\bibitem[BCT22]{barg_construction_2022}
Alexander Barg, Zitan Chen, and Itzhak Tamo.
\newblock A construction of maximally recoverable codes.
\newblock {\em Designs, Codes and Cryptography}, 90(4):939--945, 2022.
\newblock Publisher: Springer.

\bibitem[BHH12]{Blaum2012PartialMDSCA}
Mario Blaum, James~Lee Hafner, and Steven Hetzler.
\newblock Partial-mds codes and their application to raid type of
  architectures.
\newblock {\em IEEE Transactions on Information Theory}, 59:4510--4519, 2012.

\bibitem[BPSY16]{blaum_construction_2016}
Mario Blaum, James~S Plank, Moshe Schwartz, and Eitan Yaakobi.
\newblock Construction of partial {MDS} and sector-disk codes with two global
  parity symbols.
\newblock {\em IEEE Transactions on Information Theory}, 62(5):2673--2681,
  2016.
\newblock Publisher: IEEE.

\bibitem[CHL07]{Chen2007OnTM}
Minghua Chen, Cheng Huang, and Jin Li.
\newblock On the maximally recoverable property for multi-protection group
  codes.
\newblock {\em 2007 IEEE International Symposium on Information Theory}, pages
  486--490, 2007.

\bibitem[CMST21]{cai_construction_2021}
Han Cai, Ying Miao, Moshe Schwartz, and Xiaohu Tang.
\newblock A construction of maximally recoverable codes with order-optimal
  field size.
\newblock {\em IEEE Transactions on Information Theory}, 68(1):204--212, 2021.

\bibitem[CSYS15]{chen_sector-disk_2015}
Junyu Chen, Kenneth~W Shum, Quan Yu, and Chi~Wan Sung.
\newblock Sector-disk codes and partial {MDS} codes with up to three global
  parities.
\newblock pages 1876--1880. IEEE, 2015.

\bibitem[GG22]{gopi_improved_2022}
Sivakanth Gopi and Venkatesan Guruswami.
\newblock Improved maximally recoverable {LRCs} using skew polynomials.
\newblock {\em IEEE Transactions on Information Theory}, 2022.

\bibitem[GGY20]{gopi_maximally_2020}
Sivakanth Gopi, Venkatesan Guruswami, and Sergey Yekhanin.
\newblock Maximally recoverable {LRCs}: {A} field size lower bound and
  constructions for few heavy parities.
\newblock {\em IEEE Transactions on Information Theory}, 66(10):6066--6083,
  2020.

\bibitem[GHJY14]{gopalan_explicit_2014}
Parikshit Gopalan, Cheng Huang, Bob Jenkins, and Sergey Yekhanin.
\newblock Explicit maximally recoverable codes with locality.
\newblock {\em IEEE Transactions on Information Theory}, 60(9):5245--5256,
  2014.

\bibitem[GHSY12]{gopalan_locality_2012}
Parikshit Gopalan, Cheng Huang, Huseyin Simitci, and Sergey Yekhanin.
\newblock On the locality of codeword symbols.
\newblock {\em IEEE Transactions on Information theory}, 58(11):6925--6934,
  2012.

\bibitem[GJX20]{guruswami_constructions_2020}
Venkatesan Guruswami, Lingfei Jin, and Chaoping Xing.
\newblock Constructions of maximally recoverable local reconstruction codes via
  function fields.
\newblock {\em IEEE Transactions on Information Theory}, 66(10):6133--6143,
  2020.

\bibitem[GS21]{gabidulin_rank_2021}
Ernst~M Gabidulin and Vladimir Sidorenko.
\newblock {\em Rank codes}.
\newblock 2021.

\bibitem[GYBS18]{gabrys_constructions_2018}
Ryan Gabrys, Eitan Yaakobi, Mario Blaum, and Paul~H Siegel.
\newblock Constructions of partial {MDS} codes over small fields.
\newblock {\em IEEE Transactions on Information Theory}, 65(6):3692--3701,
  2018.
\newblock Publisher: IEEE.

\bibitem[HSX{\etalchar{+}}12]{huang_erasure_2012}
Cheng Huang, Huseyin Simitci, Yikang Xu, Aaron Ogus, Brad Calder, Parikshit
  Gopalan, Jin Li, and Sergey Yekhanin.
\newblock Erasure coding in windows azure storage.
\newblock pages 15--26, 2012.

\bibitem[HY16]{hu_new_2016}
Guangda Hu and Sergey Yekhanin.
\newblock New constructions of {SD} and {MR} codes over small finite fields.
\newblock {\em 2016 IEEE International Symposium on Information Theory}, pages
  1591--1595, 2016.

\bibitem[KG05]{gabidulin}
A.~Kshevetskiy and E.~Gabidulin.
\newblock The new construction of rank codes.
\newblock In {\em Proceedings. International Symposium on Information Theory,
  2005}, pages 2105--2108, 2005.

\bibitem[LX22]{liu_maximally_2022}
Shu Liu and Chaoping Xing.
\newblock Maximally {Recoverable} {Local} {Repairable} {Codes} from {Subspace}
  {Direct} {Sum} {Systems}.
\newblock {\em IEEE Transactions on Information Theory}, 2022.
\newblock Publisher: IEEE.

\bibitem[MP22]{martinez-penas_general_2022}
Umberto Martínez-Peñas.
\newblock A general family of {MSRD} codes and {PMDS} codes with smaller field
  sizes from extended {Moore} matrices.
\newblock {\em SIAM Journal on Discrete Mathematics}, 36(3):1868--1886, 2022.

\bibitem[MPK19]{martinez-penas_universal_2019}
Umberto Martínez-Peñas and Frank~R Kschischang.
\newblock Universal and dynamic locally repairable codes with maximal
  recoverability via sum-rank codes.
\newblock {\em IEEE Transactions on Information Theory}, 65(12):7790--7805,
  2019.
\newblock Publisher: IEEE.

\bibitem[PD12]{Papailiopoulos2012LocallyRC}
Dimitris Papailiopoulos and Alexandros~G. Dimakis.
\newblock Locally repairable codes.
\newblock {\em IEEE Transactions on Information Theory}, 60:5843--5855, 2012.

\end{thebibliography}

\end{document}